\documentstyle[12pt]{article}

\newcommand{\be}{\begin{equation}}
\newcommand{\ee}{\end{equation}}

\newcommand{\ba}{\begin{eqnarray}}
\newcommand{\ea}{\end{eqnarray}}

\newcommand{\sumN}{\sum_{i=1}^N}
\newcommand{\sumi}{\sum_i}

\newcommand{\prodi}{\prod_i}

\newcommand{\prodjnoi}{\prod_{j(\not= i)}}

\newcommand{\intinfinf}{\int\limits_{-\infty}^{\infty}}

\newcommand{\hi}{\{h_i\}}

\newcommand{\de}{\delta}
\newcommand{\ep}{\epsilon}
\newcommand{\e}{\eta}
\newcommand{\om}{\omega}
\newcommand{\r}{\rho}
\newcommand{\s}{\sigma}
\newcommand{\ta}{\tau}
\newcommand{\th}{\theta}
\newcommand{\thi}{\theta_i}
\newcommand{\D}{\Delta}
\newcommand{\Th}{\Theta}
\newcommand{\cD}{{\cal D}}
\newcommand{\cV}{{\cal V}}
\newcommand{\cF}{{\cal F}}
\newcommand{\pa}{\partial}
\newcommand{\sign}{\mbox{sign}}

\font\ninerm=cmr9

\begin{document}

\hsize36truepc\vsize51truepc
\hoffset=-.4truein\voffset=-0.5truein
\setlength{\textheight}{8.5 in}

\begin{titlepage}
\begin{center}
\hfill LPTENS-94/28\\
\hfill Oct 1994\\

\vskip 0.6 in
{\large Large $N$ Phase Transition In The Heat Kernel On The
$U(N)$ Group}
\vskip .6 in
       {\bf Vladimir A. Kazakov \\}
            {\it and \\}
       {\bf Thomas Wynter$^a$}\vskip 0.3cm

        \vskip 0.3 cm
{\it Laboratoire de Physique Th\'eorique de l'\'Ecole Normale
Superi\'eure$^b$\\
24 rue Lhomond, 75231 Paris Cedex 05, France\\}

\vskip 0.6 in

{\bf Abstract}
\end{center}
\vskip 14.5pt

{\leftskip=0.5truein\rightskip=0.5truein\noindent
{\small

The large N phase transition point is investigated in the heat kernel
on the $U(N)$ group with respect to arbitrary boundary conditions.
A simple functional relation is found relating the density
of eigenvalues of the boundary field to the saddle point shape of the
typical Young tableaux in the large $N$ limit of the character
expansion of the heat kernel. Both strong coupling and weak coupling
phases are investigated for some particular cases of the boundary holonomy.
}
\par}

\vskip 2.2in
\hrule width5cm \vskip 2pt
{\ninerm \hskip -17pt $^a$
This work is supported by funds provided by the European Community,
Human Capital and Mobility  Programme\\
$^b$ Unit\'e Propre du Centre National de la Recherche
Scientifique, associ\'ee \`a l'\'Ecole Normale Sup\'erieure et \`a
l'Universit\'e de Paris-Sud.}

\newpage

\end{titlepage}
\setlength{\baselineskip}{1.5\baselineskip}

\section{Introduction}
Heat kernels on classical groups are widely used in mathematical
physics, especially in quantum field theory \cite{Dowker} \cite{Menotti}.
For example it serves as
a building block (disc partition function with an arbitrary boudary
holonomy) for calculations of the partition function of
Yang-Mills theory on 2D manifolds with various topologies
\cite{Migdal75} \cite{Rusakov} \cite{Gross}
\cite{GrossTay}. The
large $N$ limit of the heat kernel is of special interest since it is
related to the so called planar limit in some quantum field theories,
QCD$_2$ being the most interesting among them \cite{KazKos1}
\cite{Kaz} \cite{KazKos2} \cite{Kos}.
This limit not only can simplify the analytical investigation
of a theory but can also give rise to new qualitative phenomena, like
large $N$ phase transitions \cite{DougKaz}.

Our task in this paper is to calculate, as explicitly as possible, the
heat kernel for the $U(N)$ group (as an example) in the large $N$ limit.
The method will be closely related to the recent calculation of the
partition function of Yang Mills theory on the two dimensional sphere
\cite{DougKaz}.

To give a physical taste to the heat kernel let us derive the heat kernel
as the partition function of a simple quantum mechanical model;
the one dimensional principal chiral field with initial and final
holonomies specified . The functional integral representation of the
partition function is
\be
Z(T,\th_0,\th_T)=\int\bigl[\cD g(t)\bigr]_H
e^{Ntr\int_0^T dt\pa_t  g^{-1}\pa_t  g},
\label{chiralZ}
\ee
where $[{\cal D}g(t)]_H$ is the Haar measure on the unitary group $U(N)$,
and the holonomies at the end points of the time interval are
$g(0)=e^{i\th_0}$  and $g(T)=e^{i\th_T}$ (with $e^{i\th_0}$ and
$e^{i\th_T}$ the eigenvalues of $g(0)$ and $g(T)$).
Due to the invariance of the Haar
measure at each moment of time, we have, after the introduction of a
new variable (connection) $A(t)=i g^{-1}\pa_t  g(t)$ the
following representation of $Z(T,\th_0,\th_T)$:
\be
Z(T,\th_0,\th_T)=\int\cD A(t)e^{-Ntr\int_0^Tdt A(t)^2}
\,\,\de\Bigl(\bigl[Te^{i\int_0^TdtA(t)}\bigr]e^{i\th_0},e^{i\th_T}\Bigr).
\label{chiral2Z}
\ee
Using the character expansion of the
group $\de$-function: $\de(U,U')=\sum_R \chi_R(U) \chi_R(U')^*$, and
the fact that
the character is the trace of the matrix element in a given
representation R:
$\chi_R(T\exp(i\int Adt))=tr_R[T\exp(i\int A^a\ta^a_Rdt)]$
($\ta^a_R$
is the $a$th generator of $U(N)$ in the $R$th irreducible representation),
we arrive after a simple gaussian integration in $A(t)$ (independant at
each moment $t$) at the conclusion that
\be
Z(T,\th_0,\th_T)=\sum_R e^{-T C_2(R)}\chi_R(e^{i\th_0})\chi_R(e^{i\th_T})
\label{heatkerndouble}
\ee
where $C_2(R)$ is the second Casimir of the group $U(N)$ in the
representation $R$.
This is also equivalent to the partition function of two dimensional
QCD on the cylinder with the
boundary holonomies specified at either end of the cylinder by the two
distributions $e^{i\th_0}$ and $e^{i\th_T}$. For the rest of this
article we restrict our attention to the case where one of the
holonomies is the unit matrix (two dimensional QCD on the disc),
and the heat kernel takes the form
\be
Z(T,\th)=\sum_R d_R e^{-T C_2(R)}\chi_R(e^{i\th}).
\label{heatkern}
\ee

In section(2) we derive an explicit formula
for the partition function for the heat kernel with fixed
boundary field. The solution is written directly in terms of the
eigenvalues of the boundary field. The calculation is valid only in
the weak coupling phase. In section(3) we find a simple
inversion relation between the density of eigenvalues of the boundary
field and the density of heighest weights labeling the saddle point
Young tableau. In some important particular cases of boundary
holonomy, such as the semicircle distribution, it
is shown to be explicitly solvable.
{}From this it is then trivial to calculate the strong
coupling transition point. We then discuss in section(4) how to
calculate the partition function in the strong coupling phase. An
alternative approach is
discussed in the appendix. The appendix also contains an
explicit solution for the heat kernel action illustrating the
procedure developped in section(3) for a nontrivial boundary field
(the density of the $\th$'s being given by an inverted
semicircle distribution $\s(\th)={1\over\pi\sqrt{g^2-\th^2}}$).

\section{Weak coupling phase}
To study the action in the large $N$ limit we proceed as in \cite{DougKaz}.
The sum over representations $R$ for the group $U(N)$ is the sum over all
Young tableaux labeled by the components of heighest weight
${n_1,n_2,...,n_N}$. The $n_i$ are integers, and the sum over
representations is given by summing over all $\{n_i\}$ satisfying the
inequality:
\be
\infty\geq n_1 \geq n_2 \geq .... \geq n_N \geq -\infty.
\label{nconst}
\ee
With this labeling for the representation, the explicit expressions for
the dimension, second Casimir and character are:
\ba
& d_R=\prod_{i>j}(1-{n_i-n_j\over i-j}),\\
& C_2(R)=\sumN n_i(n_i-2i+N+1),\\
& \chi_R(e^{i\theta})={det\hskip -1 pt _{_{k,l}}(e^{i(n_k-k+N)\th_l})
                        \over \D(e^{i\th})},
\label{ndefns}
\ea
where the determinant is of the matrix whose $(k,l)$ element is
$e^{i(n_k-k+N)\theta_l}$ and $\Delta(e^{i\theta})$ is the
Vandermondt determinant $\D(x)=\prod_{i>j}(x_i-x_j)$.

It is convenient to introduce a new set of labels, $\hi$ defined by
\be
h_i=-n_i+i-(N+1)/2,
\label{h}
\ee
which satisfy the inequality
\be
h_{i+1}> h_i.
\label{hconst}
\ee
The heat kernel is then given by
\be
Z=\sum_{\{h_i\}_{constrained}}\D(h)
                    e^{-{T\over N}\sum_{i=1}^Nh_i^2}
det\hskip -1 pt _{_{k,l}}(e^{-ih_k\th_l})\,f(\th)
\label{Zorig}
\ee
where
\be
f(\th)={const\over \D(e^{i\th})}\,\,
        e^{-{T\over 12}(N^2-1)+i{N-1\over 2}\sumN\thi},
\label{fdefn}
\ee
and the sum over the $\{h_i\}$ is an ordered sum subject to the
inequality (\ref{hconst}). The determinant consists of $N!$
terms of alternating sign. By interchanging the $\{h_i\}$, each of
these terms can be made equal to $e^{(-i\sum_{i=1}^Nh_i\theta_i)}$,
the ordered sum becomes unconstrained, (each of the $N!$ terms
contributing a different segment of the sum over all $\{h_i\}$) and
the partition function becomes:
\be
Z=\sum_{\{h_i\}}\D(h)\,\,
            e^{-\sumN({T\over N}h_i^2+ih_i\thi)}
             \,\,f(\th).
\label{Zinter}
\ee
In the large $N$ limit we redefine the $\hi$, dividing them by $N$,
and then treat them as continuum
variables so that the multiple sum becomes a multiple integral:
\be
Z=\intinfinf\biggl(\prodi dh_i\biggr)\,\,\D(h)\,\,
                   e^{-N\sumi Th_i^2+ih_i\thi}\,\,\,f(\th).
\label{Zcont}
\ee
It is important to realise that by changing the sum into an integral
we may have lost the inequality (\ref{hconst}). The expression for $Z$
we now obtain is therefore only valid in the weak coupling phase where
(\ref{hconst}) is satisfied.
By shifting the $\hi$ we can complete the square and do the
gaussian integral:
\ba
Z=&\intinfinf\biggl(\prodi dh_i\biggr)\,\,\D(h-{i\over 2T}\th)\,\,
     e^{-N\sumi Th_i^2+{1\over 4T}\thi^2}\,\,\,f(\th)\\
  =&\D(i\th)e^{-{N\over 4T}\sumi\thi^2}\,\,\,T^{-N^2/2}f(\th).
\label{Zwk}
\ea
The last step is accomplished by noting that the answer is a
completely antisymmetric polynomial of the $\theta_i$ of the same
degree as the Vandermondt determinant,
ie., it is the Vandermondt determinant up to a constant.  We have
thus restored the partition function in the weak coupling phase \cite{Menotti}.
We will see in the next section that the weak coupling solution will
be valid only for small enough $T$, due to the constraint (\ref{hconst}).

\section{Strong Coupling Transition Point}
It is expected that in the large $N$ limit the integral above is dominated
by the contribution from a single set of $\{h_i\}$ ie., a single
representation. In terms of the continuum variables, this corresponds
to a density $\r(h)={1\over N}{\pa i\over \pa h}$ for the distribution of
$h$'s.
As is discussed in \cite{DougKaz}, for this distribution to correspond
to an allowed Young tableau $\r(h)$
must satisfy the constraint $\r(h)\leq 1$, following directly from
(\ref{hconst}).  The point at which $\r(h)=1$, for some $h$,
corresponds to the strong coupling transition point beyond
which the weak coupling solution is no longer valid, and the strong
coupling methods discussed in the following section have to be used.
The rest of this section is entirely devoted
to finding the distribution $\r(h)$ corresponding to the weak coupling
solution, to allow us to test for this
condition.

To find the density we calculate the resolvent, $H(h)$, which is
defined to be:
\be
H(h)=\biggl\langle{1\over N}\sumN{1\over h-h'_i}\biggr\rangle_{h'},
\label{discresolv}
\ee
where h lies in the complex plane and the average is taken with
respect to $h'$. In the large $N$ continuum limit this is given as the
integral along the support of $\r(h)$
\be
H(h)=\int dh{\r(h')\over h-h'},
\label{contresolv}
\ee
The discontinuity across the cut of the resolvent gives $2\pi i$ times
the density, $\r(h)$.

To proceed we take the Fourier transform of each term in the sum
(\ref{discresolv}):
\be
H(h)={1\over N}\sumN {i\over 2}\intinfinf dk\,\sign(k)e^{-ikh}
           \Bigl\langle e^{ih'_ik} \Bigr\rangle_{h'}
\label{FourH}
\ee
The expectation value, $\langle e^{ih'_ik} \rangle_{h'}$, can
be calculated from the formulae (\ref{Zcont}) and (\ref{Zwk}) for $Z$
by realising that
the term $e^{ih_ik}$ can be absorbed into the definition of the $\th$.
We therefore define a new distribution of $\th$'s which is identical to the old
distribution except for the $i$th $\th$ which is shifted by $-k/N$;
\be
\tilde\th_j=\th_j-{k\over N}\de_{ij}.
\label{thetatild}
\ee
This allows us immediately to write down the result
\ba
\Bigl\langle e^{ih'_ik} \Bigr\rangle_{h'}=&{\D(\tilde\th)
                                          e^{-{N\over4T}\sum_j\tilde\th_j^2}
                                          \over
                                          \D(\th)
                                          e^{-{N\over4T}\sum_j\th_j^2}}
                    &=\prodjnoi\Bigl(1-{k\over N(\thi-\th_j)}\Bigr)
                       e^{-{1\over 4T}\bigl({k^2\over N}-2\thi k\bigr)}
\label{expexpihk}
\ea
Substituting this into equation(\ref{FourH}) we notice that the sum
can be replaced by a contour integration in $\th$ (encircling all
the $\th$ eigenvalues) if we also unrestrict the product, so that
the product is over all $j$:
\be
H(h)=\int_{-\infty}^{\infty}\,dk{\sign(k)e^{-ikh}\over 2k}
\,\,{1\over 2\pi}\oint\,d\th\prod_{j=1}^N\Bigl(1-{k\over N(\th-\th_j)}\Bigr)
e^{-{1\over 4T}\bigl({k^2\over N}-2\th k\bigr)}.
\label{contourprod}
\ee
We can then exponentiate the unrestricted
product and keep only the lowest order in $1/N$ to arrive at
\be
H(h)={1\over 2\pi}\oint\,d\th\int_0^{\infty}\,dk\,\,\,
      \bigl(e^{-{k^2\over 4NT}}\bigr)\,\,\,
      {-\cos k(h+i{\th\over 2T}-i\Th(\th))\over k}
\label{Matthias}
\ee
where
\be
\Th(\th)=\int\,d\th'{\s(\th')\over \th-\th'}
            ={1\over N}\sum_j{1\over \th-\th_j}+O({1\over N})
\label{thetaresol}
\ee
is the resolvent for the distribution of $\th$.
The apparent divergence of the $k$ integral about $k=0$ is canceled by
the contour integration in $\th$ (we can add any non $\th$ dependant
function of $k$ into the $k$ integrand, eg.
$(e^{-{k^2\over 4NT}}){\cos(k)\over k}$, since
it will be ignored by the contour integral). The gaussian exponential
term suppresses the divergent behaviour of the cos (it's argument is complex)
and permits us to do the integral unambiguously in the large $N$
limit, the result being:
\be
H(h)={1\over 2\pi}\oint_C\,d\th \log\bigl[h+i{\th\over
2T}-i\Th(\th)\bigr]
\label{contourH}
\ee
where the contour $C$ goes anticlockwise around the
cut of the resolvent $\Th(\th)$. We would like to thank Matthias
Staudacher for simplifying this derivation.

In the appendix we show how to find
an analagous formula (the
roles of $H(h)$ and $\Th(\th)$ are interchanged) using either Migdal's method
\cite{Migdal} or some results developped by Matytsin \cite{Matyt}.

It is more convenient to study the derivative of $H(h)$:
\be
H'(h)={1\over 2\pi}\oint_C\,d\th{1\over\bigl[h+i{\th\over 2T}-i\Th(\th)\bigr]}
\label{contourH'}
\ee
In this case
the cut point(s) of the logarithm becomes a pole(s) and the contour
can be inflated out from
the cut to catch both the pole(s) and also a contribution from infinity:
\be
H'(h)=2T-\sum_{zeros}{i\over\bigl[{i\over 2T}-i\Th'(\th(h))\bigr]}
\label{ResH'}
\ee
where the $\th(h)$ satisfy
\be
h+i{\th(h)\over 2T}-i\Th(\th(h))=0
\label{thetah}
\ee
and the sum over the zeros means the sum over all the $\th(h)$
which satisfy equation(\ref{thetah})
Taking the derivative with respect to $h$ of equation(\ref{thetah})
gives an identity which allows us to reduce equation(\ref{ResH'}) to
\be
H'(h)=2T+i\sum_{zeros}\th'(h)
\label{H'}
\ee
Integrating up with respect to $h$ we arrive at the very simple
formula
\be
H(h)=2Th+i\sum_{zeros}\th(h)
\label{H}
\ee
Some general comments are now in order. Firstly, we only need to solve
the inversion equation(\ref{thetah}) in some neighbourhood of $h$, as analytic
continuation then defines $H(h)$ over the whole complex
plane. The cutpoints are completely specified, although the actual
path of the cuts between the cutpoints maybe ambiguous, in which case
they are chosen to lie on the real axis. For densities $\s(\th)$ which
are non-singular, ie. finite on all of their support, we can choose the
neighbourhood of $h$ to be far from the origin so that there is only
one solution of equation(\ref{thetah}). In this case we have
\be
H(h)=2Th+i\th(h)=2Ti\Th(\th(h))
\label{onezeroH}
\ee
where the last equality follows directly from (\ref{thetah}).
Studying the large $h$ behaviour of the right hand side we find that
$H(h)$ behaves as ${1\over h}$ for large $h$ as is required.
For this reason we have neglected the possible constant term
in equation(\ref{H}) coming from the integration.

We are now in a position to find the strong coupling transition point,
ie. to test the inequality $\r(h)\leq 1$. If the distribution of $h$'s
reaches it's maximum at $h=0$ we can derive a very simple expression
for the strong coupling transition point. Setting $h=0$ in
equation(\ref{onezeroH}) we have at the strong coupling transition
point $\Th(\th(0)))=\pm\pi/2T$, which implies (from
equation(\ref{thetah})) that $\th(0)=\pm\pi$ which leads in turn to
the simple expression for the strong coupling transition point:
\be
T_c={\pi\over 2\Th(\pi)}
\label{matcrit}
\ee
We would like to thank Andrei Matytsin for this comment \cite{Matcomment}.

\subsection{An example:  The semicircle distribution}
To illustrate the method discussed above we present the simplest
example, the case where the spectrum of $\th$'s is given by a semicircle
distribution:
\be
\s(\th)={2\over \pi c}\sqrt{c-\th^2}\,\,\,\,\,\,\mbox{and}\,\,\,\,\,
\Th(\th)={2\over c}\bigl(\th-\sqrt{\th^2-c}\bigr)
\label{semicirctheta}
\ee
Inverting equation(\ref{thetah}) gives
\be
\th(h)=2ih\bigl(T-{1\over c_h})+{2i\over c_h}\sqrt{h^2-c_h}
\label{thetahsemicirc}
\ee
and leads to a semicircle distribution for the $h$'s
\be
\r(h)={2\over \pi c_h}\sqrt{c_h-h^2}
\label{semicirch}
\ee
where the constant $c_h$ is related to $c$ by
\be
c_h={2\over T}-{c\over 4T^2}.
\label{ch}
\ee
The strong coupling transition point occurs for the value of $T$ at which
$\r(h=0)=1$, giving:
\be
c_h={4\over \pi^2} \Longrightarrow T={\pi\over {4\over c}(\pi-\sqrt{\pi^2-c})}
\label{chcrit}
\ee
This can also be retrieved directly from equation(\ref{matcrit}).
As a check we set $c=0$ so that $M(\th)$ is the unit
matrix to give the partition function of $QCD_2$ on the sphere and
obtain the critical point discovered in \cite{DougKaz} which is
$2T_{crit}=\pi^2$, where $2T$ is the area.

Another exactly solvable distribution $\s(\th)={1\over
\pi\sqrt{g^2-\th^2}}$ is studied in the appendix. This
distribution is singular at the end points and provides an
illustration of the case where there are several zeros to
equation(\ref{thetah}).

\section{The Strong Coupling Phase}
In this section we present the analogue of the method described in
\cite{DougKaz} for calculating the distribution of $h$'s and the free
energy in the strong coupling phase. In \cite{DougKaz} the procedure
was to start with the saddle point equation and look for the density
$\r(h)$ which satisfies it subject to the constraint that for a finite
interval $[-b,b]$ $\r(h)$ takes on it's maximal value, ie.  $\r(h)=1$
for $h\ep [-b,b]$. The
actual value of $b$ is then found from consistency
arguments.

Although we derived our results from a completely different direction
we still have a saddle point equation. It is the real part of
equation (\ref{H}) ie.
\be
P\int\,dh'{\r(h)\over h-h'}=\Re H(h)_{\mbox{\tiny weak}}
                              =2Th-\Im\sum_{zeros}\th(h)
\label{saddlept}
\ee
The procedure is thus to solve the problem in the weak coupling phase
by the inversion described above, and obtain $H(h)_{\mbox{\tiny weak}}$
as an explicit
expression of $h$, take the real part of this equation and solve
(\ref{saddlept})
for $\r(h)$ subject to the constraint that $\r(h)\leq 1$.
To do this we follow exactly the method discussed in \cite{DougKaz},
which we repeat here almost verbatim.
For simplicity's
sake, we assume here that the distribution of $h$'s is symmetric
and that there is only one interval, $[-b,b]$, for which $\r(h)=1$.
\be
\r(h)=\cases{1 &for $h\ep[-b,b]$;\cr
      \tilde\r(h) &elsewhere.\cr }
\label{newrho}
\ee
Substituting into the real part of (\ref{saddlept}) we get
\be
\Re H(h)_{\mbox{\tiny weak}}-\log{h-b\over h+b}=
                             P\int dh{\tilde\r(h')\over h-h'}.
\label{saddlepoint2}
\ee
We define the resolvent for the distribution $\tilde\r(h)$
\be
\tilde H(h)=\int{dh'\tilde\r(h')\over h-h'}
\label{tilderesolv}
\ee
where the integral runs along the support of $\tilde\r(h)$, $[-a,-b]$
and $[b,a]$ (with $a$ and $b$ to be found later from the large $h$
behaviour of $\tilde H(h)$), and generate the full analytic function
$\tilde H(h)$ from  $\Re H(h)_{\mbox{\tiny weak}}$ by performing the
contour integral
\be
\tilde H(h)=-{1\over 2\pi i}\sqrt{(a^2-h^2)(b^2-h^2)}\oint\,ds
{\Re H(s)_{\mbox{\tiny weak}}-log{s-b\over s+b}
                             \over (s-h)\sqrt{(a^2-h^2)(b^2-h^2)}}
\label{cauchy}
\ee
where the contour encircles the cut of the square root but does not
enclose any other cuts or singularities of the integrand.
By inflating the contour we catch the pole at $h=s$, the cut of the
$\log$ and any singularities of the analytic continuation of
$\Re H(h)_{\mbox{\tiny weak}}$. The full resolvent
$H(h)_{\mbox{\tiny strong}}$  for the
strong coupling phase involves adding to $\tilde H(h)$ the contribution
from the constant part of the full density $\r(h)$. This precisely
cancels out the $\log$ term generated by the pole at $h=s$.
We thus arrive at the expression for the resolvent for the strong
coupling phase
$$H(h)_{\mbox{\tiny strong}}=\Re H(h)_{\mbox{\tiny weak}}-
 \sqrt{(a^2-h^2)(b^2-h^2)}\int_{-b}^b\,ds
   {1\over (s-h)\sqrt{(a^2-h^2)(b^2-h^2)}}$$
\be
\hskip 50pt +(\mbox{ contribution from singularities of } \Re
H(h)_{\mbox{\tiny weak}}).
\label{strongresolv}
\ee
The last two terms give plus or minus $i\pi$ times the full
density $\r(h)$. As is discussed in \cite{DougKaz} the second term
can be written entirely in terms of the complete elliptic integral
of the third kind $\Pi[\phi,x]$:
\be
{1\over\pi}{b-a\over b+a}\sqrt{{(a+h)(b+h)\over (a-h)(b-h)}}
            \Pi\biggl[{2b\over a+b}{h-a\over h+b},{2\sqrt{ab}\over a+b}\biggr]
\label{eliptint}
\ee
The coefficients $a$, and $b$ are found
by ensuring that for large $|h|$  $H(h)_{\mbox{\tiny strong}}={1\over
h}+O({1\over h^2})$. The large $h$ limit of the elliptic integral is
discussed in \cite{DougKaz}.

For the semicircle distribution given as an example at the end of
section (3) we have
$\Re H(h)_{\mbox{\tiny weak}}={2\over c_h}h$. All the analysis
conducted in \cite{DougKaz} for the strong coupling phase is
as before if we make the replacement in the notation of \cite{DougKaz} of
$A/2\rightarrow T/(1-{c\over 2T})$.

\section{Appendix }
\subsection{An alternative approach}
Here we present an alternative method to finding the
results of section(2). The charcter for $U(N)$ is, up to a factor of
$i$ and a couple of trivial Vandermondt factors, the Itzykson Zuber determinant
\be
\chi_R(e^{-i\th})=\D(h)I(h,-i\th){\D(i\th)\over\D(e^{i\th})}
        \prod_i e^{i{N-1\over 2}\th_i}
\label{charequIZ}
\ee
\be
\mbox{where}\,\,\,\,\,\,\,\,
I(\phi,\psi)=\int\,dU_He^{N\,tr(U\phi U^{\dagger}\psi)}
\rightarrow e^{N^2F_0[\r,\s]}\,\,\,\,\mbox{as}\,\,\,\,\,N\rightarrow\infty .
\label{ItzykZub}
\ee
Migdal \cite{Migdal}, in the context of induced QCD, and more recently
Matytsin \cite{Matyt} derived some useful results for the
large $N$ limit of the Itzykson Zuber integral:
$I\bigl[\r(\phi),\s(\psi)\bigr]=e^{N^2F_0[\r,\s]}$.
We start from a useful formula which comes from taking the
large $N$ limit of
${tr\over N}\bigl({\partial\over\partial \phi}\bigr)^q I(\phi,\psi)$.
{}From Matytsin's paper \cite{Matyt} we quote the result (with the minor
change of an extra factor of $-i$)
$$
\hskip -50pt \int\,d\th'\s(\th')(-i\th')^q=
{1\over 2\pi i}{1\over q+1}\int\,dh'\biggl[\Bigl(
 {\partial\over\partial h}{\de F_0[\r,\s]\over\de\r(h')}+\cV(h')+i\pi\r(h')
                                             \Bigl)^{q+1}
$$
\be
\hskip 190pt -\Bigl(
 {\partial\over\partial h}{\de F_0[\r,\s]\over\de\r(h')}+\cV(h')-i\pi\r(h')
                                             \Bigl)^{q+1}\biggr] .\\
\label{Matyt}
\ee
The integral runs along the support of $\r(h)$.
We recognise $\cV(h)\mp i\r(h)$ as the value of the resolvent $H(h)$
above and below the cut and the equation simplifies to
\be
\int\,d\th'\s(\th')(\th')^q={-1\over 2\pi}{1\over q+1}\oint_C\,dh
           \bigl[i\cF(h)+iH(h)\bigr]^{q+1}
\label{contourMatyt}
\ee
where for convenience we have defined
$\cF={\partial\over\partial h}{\de F_0[\r,\s]\over\de\r(h')}$, and the
contour integration circles the cut.
Multiplying both sides of the above equation by ${1\over\th^{q+1}}$ and
summing from $q=0$ to $\infty$ we obtain
\be
\Th(\th)=
  \oint_C\,{dh\over 2\pi}\log\bigl[\th-i\cF(h)-iH(h)\bigr].
\label{contourthgen}
\ee
This is the analogue of equation(\ref{contourH}). It is an
equation derived entirely from the properties of the character
(Itzykson Zuber integral) itself.
To proceed furthur one has to replace $\cF(h)$ by its functional form derived
from the saddle point equation
evaluated for the particular action studied. It is important
to remember that $\cF$ is defined in terms of the Itzykson Zuber
integral which is related to the character by the equation(\ref{charequIZ}).
The action has first to be written in terms of the Itzykson Zuber
integral and then the saddle point equation evaluated.

For the heat kernel we rewrite the action in terms of
the Itzykson Zuber integral,
\be
Z=\int\,dh\r(h)\D^2(h)e^{-{T\over N}\sumi h_i^2}I(\r,\s)\times\D(i\th)f(\th)
\label{IZQCD2}
\ee
and then find the saddle point equation:
\be
\hskip -90pt \cF(h)=2Th-2\cV(h)
\ee
Substituting in to equation(\ref{contourthgen}) and using the identity
$H^{\pm}(h)-2\cV(h)=-H^{\mp}(h)$, where $H^{\mp}$ is the value of
$H(h)$ immediately above and below the cut, we obtain
\be
\Th(\th)=
  -\oint_C\,{dh'\over 2\pi}\log\bigl[\th-i2Th+iH(h)\bigr]
\label{contourth}
\ee
As earlier we take the derivative, expand the contour out to infinity,
picking up the poles on the way, and arrive at the equation
\be
\Th(\th)={\th\over 2T}-i\sum_{zeros}h(\th)
\label{Th}
\ee
where $h(\th)$ satisfies
\be
\th-i2Th(\th)+iH(h(\th))=0,
\label{htheta}
\ee
and the sum over zeros is the sum over all the solutions to
equation(\ref{htheta}). We have thus found the analogue of equations
(\ref{H}) and(\ref{thetah}). In the case where there are single
solutions to equations (\ref{thetah}) and (\ref{htheta}), we see that
equations (\ref{Th}) and (\ref{htheta}) are identical, respectively, to
equations (\ref{thetah}) and (\ref{onezeroH}).

\subsection{An example: The inverted semicircle distribution}
We present here the exact solution for a nontrivial density.
The spectrum of eigenvalues is
\be
\th_k=g\cos\bigl({\pi k\over N}\bigr)
\label{cosdensity}
\ee
and the corresponding density and resolvent
\be
\s(\th)={1\over\pi\sqrt{g^2-\th^2}}\,\,\,\,\,\,\,\mbox{and}\,\,\,\,\,\,\,\,
\Th(\th)={1\over\sqrt{\th^2-g^2}},
\label{quartdensity}
\ee
This density was used in the large $N$ limit of the $2D$ chiral model
studied in \cite{FatKazWieg}, where it corresponded to an external
magnetic field precisely tuned to excite the full
spectrum of mass states of the theory. The heat kernel action
investigated in this article corresponds to the 1D chiral model in an
external magneitic field.
Following the procedure of section(3), we solve
equation(\ref{thetah}). For convenience we introduce the rescaled
variables;
$x={\th\over\sqrt{2T}}$, $\e=h\sqrt{2T}$ and $b^2={g^2\over 2T}$;
in terms of which the equation is
\be
\e+ix-{i\over\sqrt{x^2-b^2}}=0
\label{inversequn}
\ee
We square up the square root and arrive at the quartic
\be
x^4-2i\e x^3-(\e^2+b^2)x^2+2bi\e x+(b^2\e^2-1)=0,
\label{quartic}
\ee

Although the quartic equation has 4 solutions only three
satisfy the equation (\ref{inversequn}). This can be intuitively
understood by looking at  $\e$ far from the origin. In this case there
is one solution in which $x$ is far from the origin ($x\approx-i\e$)
and the only two other possible solutions are the ones caught in the
singularities of the end points of the square root. The final
expression for the resolvent $H(h)$ involves summing up all three of
these solutions. Using a trivial identity $x_1+x_2+x_3+x_4=2b^2i\e$
for the roots of a quartic allows us to express $H(h)$ in terms of the
single quartic solution (single sheet), $x_4$, that does not satisfy
the inversion equation(\ref{inversequn}). Using standard formulae for
the roots of a quartic, the solution for the resolvent can be
expressed as
\be
H(h)=-\sqrt{{T\over 2}}\bigl(\e+i(\sqrt{z_1}+\sqrt{z_2}-\sqrt{z_3})\bigr)
\label{Hquart}
\ee
with $z_1$, $z_2$ and $z_3$ defined by
\ba
&z_1={2\over 3}b^2-{1\over 3}\e^2+{1\over 3}\bigl(F^++F^-\bigr)\\
&z_2={2\over 3}b^2-{1\over 3}\e^2+{1\over 3}\bigl(\om F^++\om^2F^-\bigr)\\
&z_3={2\over 3}b^2-{1\over 3}\e^2+{1\over 3}\bigl(\om^2F^++\om F^-\bigr)
\label{cubsolns}
\ea
where $\om={1\over 2}(-1+i\sqrt{3})$ is the cube root of $1$, and
$F^{\pm}$ are given by
\be
F^{\pm}=-\root3\of{k^3-18k+54b^2\pm\sqrt{(k^3-18k+54b^2)^2-(k^2-12)^3}}
\,\,\,\,\,\,\mbox{with}\,\,\,\,\,\,k=\e^2+b^2
\label{Fpm}
\ee
All the cut structure resides in the square root term inside
the definition of $F^{\pm}$. The $\sqrt{z}$ terms do not contribute
cuts since for $z\approx 0$, $z\propto \e^2$, and the square root
undoes the square. Similarly the cube roots do not contribute any cuts
since, close to their zeros, the arguments are proportional to $k^3$ and
the root undoes the cube. The zeros of the square root in
equation(\ref{Fpm}) and thus the cutpoints for the quartic are
\ba
\e_1=&\pm i\sqrt{b^2 -{1\over 3b^2}-(B^++B^-)}\\
\e_2=&\pm\sqrt{-b^2 +{1\over 3b^2}+(\om B^++\om^2B^-)}\\
\e_3=&\pm\sqrt{-b^2 +{1\over 3b^2}+(\om^2B^++\om B^-)}
\label{cutpoints}
\ea
where $\om={1\over 2}(-1+i\sqrt{3})$ and $B^{\pm}$ are defined by
\be
B^{\pm}=-\root 3 \of{-{1\over\beta^3}+{20\over
\beta}+8\beta\pm{8\over\beta^2}(\beta^2-1)^{{3\over 2}}}
\,\,\,\,\mbox{with}\,\,\,\,\beta=\sqrt{{27\over 4}}b^2.
\label{Bplusminus}
\ee
On the physical sheet (the sheet of the quartic corresponding to the
resolvent) there is the single pair of cut points $\pm
\e_3$. The other cut points connect
together, in pairs, the three other non-physical sheets.
The cut structure for $H(h)$ is therefore given by a
single cut or single interval for the support of the density running
from $-\e_3$ to $\e_3$.

For $b^4\leq{4\over 27}$ the support of the
density lies entirely along the real axis, but for $b^4>{4\over27}$
the end points move out into the complex plane. This is a signal that
there is no longer a single Young tableau contributing to the sum.
Probably it can be interpreted as the
manifestation of the fact that a ``density'' of terms of alternating
sign contributes in the large $N$ limit rather than a single term.

There are two cases where
everything simplifies (the simplification reduces the inversion of
equation(\ref{inversequn}) to the solving of a quadratic). The first is
when $b=0$, which corresponds
to all the eigenvalues lying at the origin, ie. when $M(\th)$ is the unit
matrix. In this case both the cube root and the square root around the
$z_i$ can be taken explicitly and a semicircle distribution recovered
as before. The other
more useful case is when h=0. Again the cube roots and square roots
around the $z_i$ can be taken explicitly and we get a simple
expression for the discontinuity across the cut at the origin ie
\ba
\r(h=0)&={\sqrt{2T}\over \pi}\sqrt{\sqrt{{b^4\over 4}+1}-{b^2\over 2}}\\
       &={1\over \pi}\sqrt{\sqrt{{g^4\over 4}+4T^2}-{g^2\over 2}}.
\label{critpt}
\ea
Since the density for the $h$'s
reaches it's maximum at the origin we have found the critical point
for the distribution ie. $\r(h)_{max}=1$ at $T={\pi\over 2}\sqrt{\pi^2-g^2}$.
This result could also have been obtained directly from
equation(\ref{matcrit}).

\vskip 14.5pt\hskip -18pt{\Large \bf Note}

Since the completion of this work it has come to our attention that D.
Gross and A. Matytsin \cite{GrossMat} have also been studying the
large $N$ limit of equation(\ref{heatkerndouble}) and have found,
amongst other things, a similar simple inversion relation for the
density of heighest weights for the case where $g(0)= g(T)^{\dagger}$.

\vskip 14.5pt\hskip -18pt{\Large \bf Acknowledgements}

We would like to thank Andrei Matytsin for the simple expression for
the strong coupling transition point. We are also grateful to Matthias
Staudacher, Edouard Brezin and Michael Douglas for many fruitful
discussions.

\setlength{\baselineskip}{0.666666667\baselineskip}

\end{document}